\documentclass[10pt,journal,compsoc]{IEEEtran}

\usepackage{setspace} 
\usepackage{latexsym, graphicx, textcomp}
\usepackage{epsfig,upgreek}
\usepackage{amsfonts, amssymb, amsthm,makecell,enumerate}
\usepackage{float,color,dblfloatfix}
\usepackage{fancyhdr}
\usepackage{amsmath}
\usepackage{amsfonts}
\usepackage{mathrsfs}
\usepackage{pifont}
\usepackage{amssymb}
\usepackage{graphicx}
\usepackage{multicol}
\usepackage{cases}
\usepackage{epstopdf}
\usepackage{ltxtable, filecontents}
\usepackage{latexsym}
\usepackage{booktabs}
\usepackage{multirow}
\usepackage{booktabs}
\usepackage{threeparttable}
\usepackage{supertabular}
\usepackage[figuresright]{rotating}
\usepackage[tight,small]{subfigure}
\usepackage{tabularx}
\usepackage{cite}
\usepackage{dcolumn}
\usepackage{algorithmicx}
\usepackage[lined,boxed]{algorithm2e}
\usepackage{algpseudocode}
 \usepackage{ragged2e}
 \usepackage{cuted}
\usepackage{graphicx} 
\usepackage{diagbox}
\usepackage{array}
\usepackage{amsfonts}
\usepackage{bbm}
\usepackage{enumitem}
\usepackage{hyperref}
\usepackage{bm}
\usepackage{hyperref}

\graphicspath{{figures/}}

\begin{document}

%

\title{\huge Unleashing the Potential of Stage-Wise Decision-Making in Scheduling of Graph-Structured Tasks \\ over Mobile Vehicular Clouds}

\author{Minghui Liwang, \IEEEmembership{Member}, \IEEEmembership{IEEE}, Bingshuo Guo, Zhanxi Ma, Yuhan Su, Jian Jin, 
\\Seyyedali Hosseinalipour,~\IEEEmembership{Member}, \IEEEmembership{IEEE}, Xianbin Wang,~\IEEEmembership{Fellow}, \IEEEmembership{IEEE}, 
Huaiyu Dai,~\IEEEmembership{Fellow}, \IEEEmembership{IEEE}


\thanks{Minghui Liwang (minghuilw@xmu.edu.cn), Bingshuo Guo (guobingshuo@stu.xmu.edu.cn) and Yuhan Su (ysu@xmu.edu.cn) are with Xiamen University, China. Zhanxi Ma is with Nanjing University, China. Jian Jin (jin.jian@caict.ac.cn) is with the Research Institute of Industrial Internet of Things, China Academy of Information and Communications Technology, China. 
Seyyedali Hosseinalipour (alipour@buffalo.edu) is with University at Buffalo-SUNY, USA. Xianbin Wang (xianbin.wang@uwo.ca) is with Western University, Canada. Huaiyu Dai (hdai@ncsu.edu) is with North Carolina State University, USA.


}}

\IEEEtitleabstractindextext{
\begin{abstract}
\justifying
To effectively process high volume of data across a fleet of dynamic and distributed vehicles, it is crucial to implement resource provisioning techniques that can provide reliable, cost-effective, and timely computing services. This article explores computation-intensive task scheduling over mobile vehicular clouds (MVCs). We use undirected weighted graphs (UWGs) to model both the execution of tasks and communication patterns among vehicles in an MVC. We then study reliable and timely scheduling of UWG tasks through a novel mechanism, operating on two complementary decision-making stages: Plan A and Plan B. 
Plan A entails a proactive decision-making approach, leveraging historical statistical data for the preemptive creation of an optimal mapping ($\alpha$) between tasks and the MVC prior to practical task scheduling. In contrast, Plan B explores a real-time decision-making paradigm, functioning as a reliable contingency plan. It seeks a viable mapping ($\beta$) if $\alpha$ encounters failures during task scheduling due to the unpredictable nature of the network. Furthermore, we provide an in-depth exploration of the procedural intricacies and key contributing factors that underpin the success of our mechanism.
Additionally, we present a case study showcasing the superior performance on time efficiency and computation overhead. We further discuss a series of open directions for future research.
\end{abstract}

}

\maketitle

\IEEEpeerreviewmaketitle

\section{Introduction}

\IEEEPARstart{T}{he} recent surge and rapid expansion of Internet-of-Vehicles (IoV), coupled with the proliferation of vehicular computation-intensive applications~\cite{15}, 
can be attributed to the advancements in communication and computing technologies. 
Many modern vehicular applications/tasks rely on machine learning (ML), demanding a substantial amount of computing power for real-time data processing. Often, this exceeds the resource capacity of a single vehicle. 
To overcome these issues, mobile edge computing (MEC)~\cite{1} has emerged as an alternative approach that brings computing, communication, and storage resources closer to vehicles by deploying them at the network edge.
In the meantime, we have witnessed the integration of smart vehicles equipped with growing on-board processing power and advanced sensors into the MEC architecture. This integration has paved the way for the migration from conventional MEC to more universal vehicular computing platforms that can provide on-demand storage and computation for emerging vehicular applications~\cite{2}. 
A paramount example of such platforms is a mobile vehicular cloud (MVC), comprising nearby moving vehicles acting as collaborating servers, enabling the distribution and simultaneous processing of tasks. MVC facilitates task processing across multiple vehicles through vehicle-to-vehicle (V2V) links, eliminating the need to transfer tasks to edge servers as in conventional MEC systems.

\begin{figure*}[t!]
\centering
\includegraphics[width=1\linewidth]{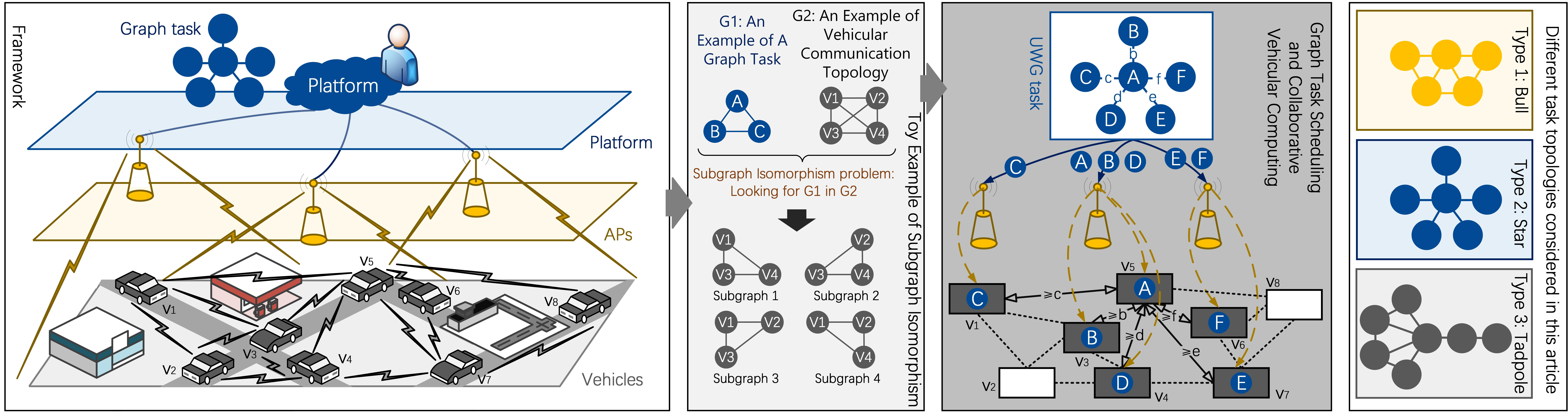}
\caption{The left-most box shows the framework for graph task scheduling over an MVC, involving vehicles as computing servers, access points, and a platform that coordinates task scheduling procedure. The second box illustrates a toy example of subgraph isomorphism problem.
The third box illustrates an example of UWG task scheduling, where the mapping between subtasks to vehicles is $\{\text{A,B,C,D,E,F}\}\leftrightarrow\{\text{v5,v3,v1,v4,v7,v6}\}$. The right-most box shows a schematic of graph task topologies (types)~\cite{6,7,8} considered in this article.}
\end{figure*}

\subsection{Preliminaries and Motivations}
This work studies the execution of computation-intensive tasks that are generated/collected periodically in an MVC. Particularly, a task is represented by an undirected weighted graph (UWG)~\cite{2,6,8} consisting of multiple subtasks, which are interconnected by weighted edges. An edge between two subtasks describes the required intermediate data exchange during task execution, while the edge weight encapsulates parameters such as the required contact duration. 
Comparing to directed acyclic graph (DAG)-structured tasks and their sequential subtask processing~\cite{4}, UWG tasks require concurrent transmission and execution of their subtasks across distinct computing servers. This makes the scheduling of UWG tasks significantly different from that of DAG tasks. Notably, the inherent parallelizability of UWG tasks necessitates stringent constraints on maintaining communication links among the computing servers, presenting pivotal challenges, particularly in MVCs with dynamic vehicles acting as servers.
Paramount examples of UWG tasks over vehicular clouds are multi-target detection, simultaneous localization and mapping~\cite{6}, and D2D-enabled federated learning, etc.

The common emphasis on graph task scheduling, involving the required resources for tasks and the distribution of subtasks across vehicles, is oriented towards the formulation of efficient \textit{on-site decision-making} mechanisms~\cite{3,6,9}. These mechanisms rely on the current network conditions, implying that task scheduling decisions are determined at the moments when tasks arrive at the system and are required to be scheduled. Nonetheless, on-site decision making faces a set of challenges. 

\begin{itemize}[leftmargin=4mm] 
 \item To schedule graph tasks over multiple servers (vehicles), finding feasible mappings between subtasks and vehicles is necessary. This requires solving the \textit{subgraph isomorphism} problem, which is NP-complete~\cite{8} (two middle subplots in Fig. 1). Additionally, network dynamics such as changing topology of vehicles, varying channel quality, and uncertainties in resource supply can complicate the problem. For instance, when two vehicles process connected subtasks, the contact duration between them should be at least equal to (or larger than) the time duration needed to finish one of the subtasks, to ensure reliable and seamless task processing.
 \item To evaluate task scheduling effectiveness, diverse metrics are utilized, encompassing the duration of task completion and energy consumption. When the optimization of multiple factors is undertaken concurrently, it gives rise to a complex problem formulation, either be a multi-objective optimization problem where achieving the pareto optimality is challenging, or a non-linear integer programming problem that is NP-hard~\cite{2}. 
\end{itemize}

As a consequence, on-site decision-making may encounter noticeable performance degradations~\cite{11,12}:
\begin{itemize}[leftmargin=4mm] 
\item \textit{Additional latency and unguaranteed task completion:} Finding feasible mappings between subtasks and vehicles can result in long delays due to the problem hardness. This further reduces the amount of time for actual resource sharing. For instance, 
during decision-making, the topology of an MVC may change, causing the obtained mapping to fail in maintaining the structure of graph tasks due to V2V links being disconnected, further result in the incompletion of tasks.

\item \textit{Excessive energy consumption:} Obtaining mappings between subtasks and vehicles is a complicated process that calls for significant computing resources, leading to high energy usage. This, in turn, can have a detrimental effect on the sustainability of the system, e.g., carbon footprint.

\item \textit{Low quality of experience:} Extended decision-making processes can adversely affect the vehicle experience within the MVC. This issue arises due to the fact that only a subset of vehicles is selected for task execution, leading to subsequent compensation, leaving others uncompensated.
\end{itemize}


\subsection{Overview and Contributions}
\textit{In this work, we aim to optimize the scheduling of UWG tasks in a MVC, with the ultimate objective of obtaining feasible mappings between subtasks and mobile vehicles, ensuring reliability, promptness, and cost-efficiency of task execution}. In light of above-mentioned challenges, we propose a novel task scheduling mechanism, which involves two complementary decision-making processes: \textit{(i)} Plan A seeks to identify the optimal mapping for future task scheduling events by analyzing historical network statistics (such a decision-making process takes place before practical task scheduling events). This mapping, referred to as $\alpha$, aims to increase the likelihood of executing future tasks with high performance. \textit{(ii)} In case $\alpha$ fails at practical/actual task scheduling events, Plan B obtains a feasible mapping ($\beta$) suitable for current network conditions (this decision-making process is triggered at the practical task scheduling events). 
Plan A contributes to the streamlining of real-time task scheduling via removing the need for on-site decision making as long as the obtained template $\alpha$ remains feasible, thereby enhancing the promptness of the overall scheduling process.
Furthermore, the incorporation of Plan B ensures the reliability of task scheduling, mitigating the risk of execution failures in real-time. \textit{This additional layer of preparedness (i.e., Plan A) differs our study from existing works through introducing a new degree of freedom in task scheduling process.}

This article provides an in-depth study of the stage-wise paradigm and reveals how it accomplishes time-efficient, cost-effective, and reliable resource provisioning in an MVC. We highlight major contributions below: 
\begin{itemize}[leftmargin=4mm] 
\item We develop a novel mechanism for scheduling graph-structured 
tasks in MVCs, involving two complementary decision-making processes. The detailed procedure (e.g., timeline), key challenges, and design considerations are discussed. 
\item We conduct a case study to demonstrate how our mechanism can be practically implemented over an MVC.
\item Through simulation results, we demonstrate that the proposed stage-wise graph task scheduling paradigm performs good on key performance metrics, including task completion time, data exchange cost, and time efficiency. Noteworthy future research directions are also discussed.
\end{itemize}

\section{Stage-Wise Decision-Making for Graph Task Scheduling over MVCs}
\subsection{Framework and Timeline}
We are interested in an IoV region (e.g., Google park), consisting of three key participants: \textit{(i)} Multiple mobile vehicles that act as computing servers; \textit{(ii)} A platform (e.g., an edge server) that generates/collects task requirements and coordinates task scheduling; \textit{(iii)} Several access points (APs, e.g., road side units), that allow vehicles to interact with the platform through vehicle-to-infrastructure (V2I) communication links. 

The platform announces graph tasks periodically, e.g., a certain number of tasks per second, while during each task scheduling event, subtasks can be allocated to vehicles for processing. Fig. 1 illustrates an example of this framework, considering a star graph task~\cite{2,8} with 6 subtasks denoted by A, B, C, D, E, and F. 
Scheduling the graph task over the MVC involves solving the \textit{subgraph isomorphism problem}, aiming to identify subgraphs within the MVC topology that mirror/replicate the structure of the UWG task. During this process, the appropriate weights are retained, ensuring, for example, that the contact duration between vehicles handling interconnected subtasks exceeds the weight associated with these subtasks. Once a specific performance goal is defined (e.g., minimizing delay and energy cost), the best isomorphic subgraph can be selected. This selection guides how subtasks are assigned across the MVC. 

To ensure an accurate portrayal of IoV, it is essential to consider key uncertainties that capture the network dynamics and thereby pose challenges to the design of task scheduling mechanisms:
\begin{itemize}[leftmargin=4mm] 
\item \textit{Fluctuant resource supply:} 
The availability of vehicles' resource supply is subject to fluctuations over time. This variability is not only influenced by the assigned graph task from the platform, but also by the vehicles' local workloads. Such fluctuations introduce uncertainty in service quality and costs. 
Additionally, forcing the prioritization of the completion of graph tasks may require additional expenses to vehicles, as their own tasks may have to wait for the release of occupied resources. These fluctuations pose risks to graph task, potentially resulting in their incomplete execution. 

\item \textit{Dynamic contact duration among vehicles:} 
When two vehicles come within their communication range, a V2V contact event occurs. Given the mobile nature of vehicles and their varying contact windows, designing a task scheduling mechanism requires safeguarding task structures. This includes facilitating data exchange through V2V links for interconnected subtasks within a graph task. Additionally, a practical mechanism needs to provide prompt scheduling decisions before the MVC topology changes.

\item \textit{Varying channel condition:} 
The wireless channel quality for both V2I and V2V links is susceptible to fluctuations over time due to various factors such as vehicle movement, obstacles, and variations in the interference of the environment. The variations in channel quality can lead to inconsistent data transmission rates, creating the potential for delays and increased energy consumption during data transmission. This variability becomes a significant concern for the successful completion of graph tasks, with the possibility of task failures if they exceed their allowable completion time. Additionally, this variability introduces overhead in interactions, manifesting as increased energy consumption for data exchange under unfavorable V2V channel conditions.
\end{itemize}

Our stage-wise task scheduling mechanism is designed to ensure reliable task completion amidst the uncertainties discussed above. It comprises two complementary decision-making plans operating on different time segments (see Fig. 2). Plan A involves analyzing historical statistics of uncertain factors, such as the historical local workload of vehicles, past information on contact duration between vehicle pairs, and the distribution of channel quality. This analysis aims to derive a mapping, denoted as $\alpha$, before practical task scheduling events. Subsequently, during each task scheduling event, Plan B assesses whether $\alpha$ is viable under the current network conditions (e.g., Events 1 and 2 in Fig. 2). In essence, Plan B examines whether the pre-determined vehicles in $\alpha$ can sustain sufficient contact duration and meet the required service quality of the task. This includes ensuring that the resources of selected vehicles and their V2V link durations can guarantee the smooth completion of their assigned subtasks. If $\alpha$ proves effective, it is executed promptly; otherwise, Plan B searches for a feasible alternative mapping, denoted as $\beta$ (e.g., Event 3), which involves selecting a set of vehicles to support task execution given the current network condition. Event 3 requires a certain period of time to obtain $\beta$, denoted by $\tau$, causing the actual service delivery to start at $t+2\Delta t+\tau$ in Fig. 2. Therefore, enhancing the feasibility of $\alpha$ and expediting the search for $\beta$ represent major challenges.

\begin{figure*}[t!]
\centering
\includegraphics[width=1\linewidth]{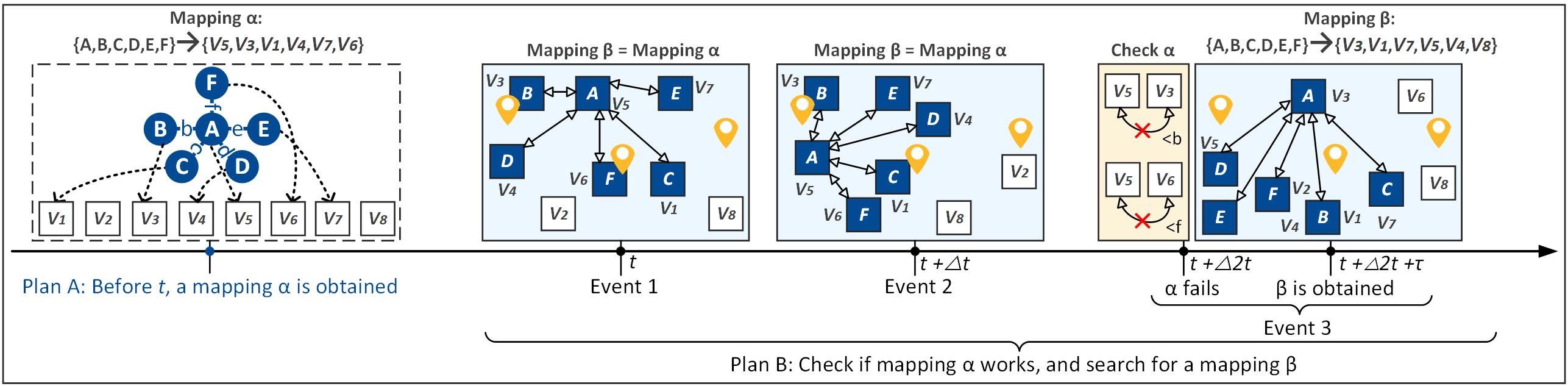}
\caption{The timeline of the designed graph task scheduling. Specifically, the three task scheduling events are independent from each other, where a graph task is generated at the beginning of each event. The considered topologies of both the graph task (star graph) and MVC are the same as that in Fig. 1. }
\end{figure*}

\subsection{Key Challenges and Considerations}
In the following, we address noteworthy considerations in the development of our stage-wise mechanism.

\noindent
\textbullet~\textit{Mapping $\alpha$: achieving the optimality and risk awareness with long-term vision}. 
This plan proactively anticipates and identifies the optimal mapping $\alpha$ by leveraging historical data and statistics related to uncertain network factors. For instance, by considering probability distributions derived from past information on uncertain factors (e.g., the V2V contact duration typically follows exponential distribution~\cite{6}), Plan A strives to determine the most favorable $\alpha$ that minimizes the expected delay and energy costs for task completion.
However, the precision of gathered information on uncertainties presents a significant challenge, as it directly influences the efficacy of $\alpha$ for future network conditions. For example, inaccuracies in statistics about the V2V contact duration can result in task execution using $\alpha$ violating task completion requirements. This is because communications among vehicles may fail to support the data exchange between connected subtasks. Utilizing mapping $\alpha$ thus entails addressing a series of risks, such as the timeout of tasks and the damage of task structure. Additionally, mapping $\alpha$ may result in the task taking longer to complete than its delay tolerance, leading to failures. Therefore, effective estimation and control of risks are crucial while striving for the optimality of mapping $\alpha$ and enhancing its applicability.

\noindent
\textbullet~\textit{Mapping $\beta$: ensuring the usability and time effectiveness with short-term vision}. Our approach for Plan B revolves around short-term objectives during each task scheduling event. This plan starts with evaluating the functionality of mapping $\alpha$ in the current network conditions, ensuring the successful completion of tasks by the vehicles associated with $\alpha$. If not, Plan B promptly addresses isomorphic subgraph search and subtask allocation optimization using the latest network information to obtain a viable mapping $\beta$. Thus, the primary concern for Plan B is the prompt acquisition of a feasible $\beta$. In particular, time-effectiveness is paramount in this process, given that the topology of the MVC, on-board resources of vehicles, and wireless channel qualities can change over time. Prolonged decision-making processes, such as spending excessive time on searching and optimizing $\beta$, can lead to task failures.

In summary, our mechanism aims to proactively establish an optimal and risk-aware mapping $\alpha$ in advance, increasing the likelihood of its successful implementation in future practical task scheduling. Additionally, it focuses on obtaining a feasible mapping $\beta$ when graph tasks are practically scheduled, serving as robust support in case $\alpha$ encounters failures. These complementary mappings contribute to prompt, cost-effective, and reliable scheduling of graph tasks in dynamic MVCs.

\section{Case Study}
This section explores a case study that examines the deployment of our proposed mechanism.

\subsection{Key Modelling}
\noindent
We consider a computation-intensive task modelled as an UWG $\bm{\mathcal{G}^{task}}$ with a subtask set, an edge set and an edge weight set. Each subtask has three attributes: tolerable completion time, data size, and the required computing resources; while each edge describes the interdependency between two subtasks~\cite{6,7}. Additionally, each edge is associated with a weight that indicates the required time for data exchange. For example, when two connected subtasks are completed on separate vehicles with completion times of 2 and 3 seconds, the edge weight is 2 seconds to ensure continuous connection during the execution. Also, the contact duration of two vehicles that handle these connected subtasks should be long enough (i.e., larger than or at least equal to the corresponding weight) to allow data transfer between subtasks.

We consider the existence of a certain number of vehicles in the MVC region, where the changes in contact duration among them reflect their dynamics \cite{6,7}. Accordingly, we model the MVC as an undirected weighted graph $\bm{\mathcal{G}^{cloud}}$ consisting of a set of vehicles, an edge set and a weight set. To capture dynamic resource supply, each vehicle in the MVC has an attribute $f$ that describes its computing capability (e.g., its CPU frequency). Also, each edge indicates the possibility of one-hop V2V communication between two vehicles. Each edge weight has two attributes: anticipated V2V contact duration $t$ between vehicles, and data exchange cost $c$ for processing of connected subtasks, e.g., energy consumed by data sharing. These attributes will be modelled as random variables to describe the network dynamics.

In this case study, we assume that the platform generates or collects graph tasks, where subtasks can be delegated to suitable vehicles for processing through a nearby AP. Note that the data transmission rates of V2I links are  represented as random variables, denoted as $r$, to capture the time-varying channel qualities. 

\subsection{How We Design the Two Plans}
Our goal in this case study is to map $\bm{\mathcal{G}^{task}}$ to $\bm{\mathcal{G}^{cloud}}$, while minimizing the task completion time (data transmission and execution delays) and the overall data exchange costs among vehicles during task processing (when two vehicles are processing connected subtasks, cost $c$ is incurred). This is achieved by formulating a cost function $F(f,c,r)$, weighting both task completion time and data exchange cost by certain coefficients, and sums them together. 


We partition the task scheduling timeline into two segments. In Plan A, our primary goal involves minimizing the expected cost function, denoted as 
$\textbf{E}[F]$. While developing Plan A, it is crucial to account for potential risks stemming from uncertainties.
Hence, Plan A involves a thorough analysis of historical statistics of the above random variables, to formalize two key risks as probabilistic constraints. The \textit{first risk} pertains to the probability of task completion time surpassing the maximum tolerable execution time for subtasks, ensuring its timely completion, based on the statistical information of $f$ and $r$. For example, the probability that the completion time fails to catch the deadline of the task. The \textit{second risk} addresses the probability of structural damage to the graph task according to the distribution of $t$, quantified by the likelihood of any vehicle pair failing to facilitate the transmission of necessary data for their assigned subtasks. For instance, the probability that the contact duration $t$ between two vehicles that are handling connected subtasks can not support the required data exchange.
Therefore, Plan A formulates an optimization problem that minimizes $\textbf{E}[F]$ while imposing control constraints on these risks within specified ranges (e.g., $\varepsilon_1$ and $\varepsilon_2$ in Fig. 3 are set as 30$\%$). 


During practical task scheduling, Plan B first checks the feasibility of mapping $\alpha$, e.g., whether the V2V connections and computing capabilities of vehicles associated with $\alpha$ can support the task structure and its tolerable completion. If $\alpha$ is unfeasible, Plan B looks for a feasible mapping $\beta$ that minimizes the practical cost function (i.e., $F$) while respecting the time and structure preservation constraints based on the current network condition (e.g., practical value of random variables). These constraints ensure the on-time task completion and data exchange between  connected subtasks to maintain the corresponding task structure.

\begin{figure}[h!t]
\centering
\includegraphics[width=0.85\linewidth]{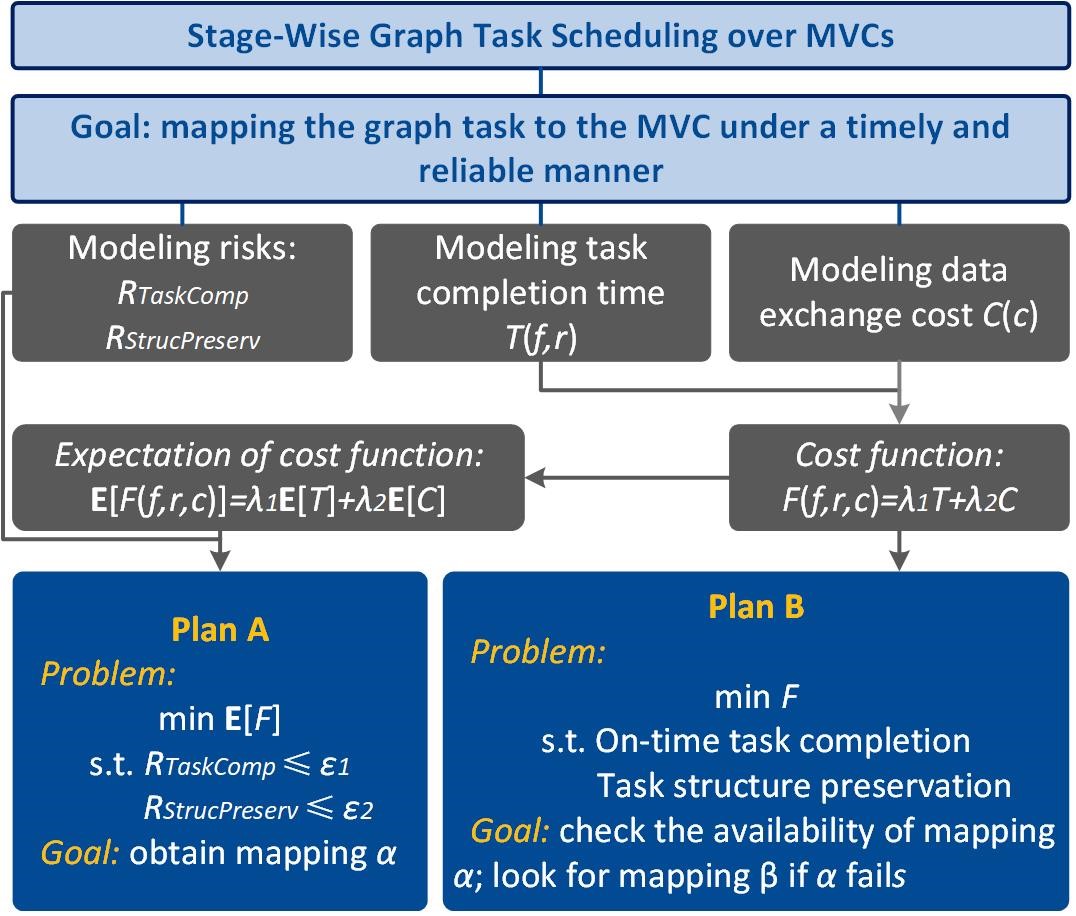}
\caption{Flow chart of our case study.}
\end{figure}

Both Plan A and Plan B center around determining the allocation and mapping vectors between subtasks and vehicles. Consequently, they confront a nonlinear integer programming problem characterized by intricate probabilistic constraints, which is NP-hard. As a result, achieving the optimal mapping within a reasonable timeframe poses a formidable challenge, especially considering that the computational time escalates exponentially with the expanding vehicular density and the intricate MVC/task topology.
To tackle this issue and optimize the subgraph search while considering probabilistic constraints, Plan A divides the problem into two subproblems. The first subproblem concerns finding feasible mappings, for which we are inspired by \cite{13}, and accomplish a risk-aware mapping search algorithm, guaranteeing that all feasible mappings adhere to reasonable risks. 
In the second subproblem, we select the optimal mapping, $\alpha$, by identifying the one with the lowest expected cost function value among all possible mappings. In practical task scheduling events, Plan B verifies if $\alpha$ is compatible with the current network. If necessary, it searches for a mapping $\beta$ using a similar algorithm to the one used in Plan A (since the time consumed in obtaining mappings in plan A is acceptable). Fig. 3 and Fig. 4 show the flow chart, and our design logic in this case study.


\begin{figure}[h!t]
\centering
\includegraphics[width=1\linewidth]{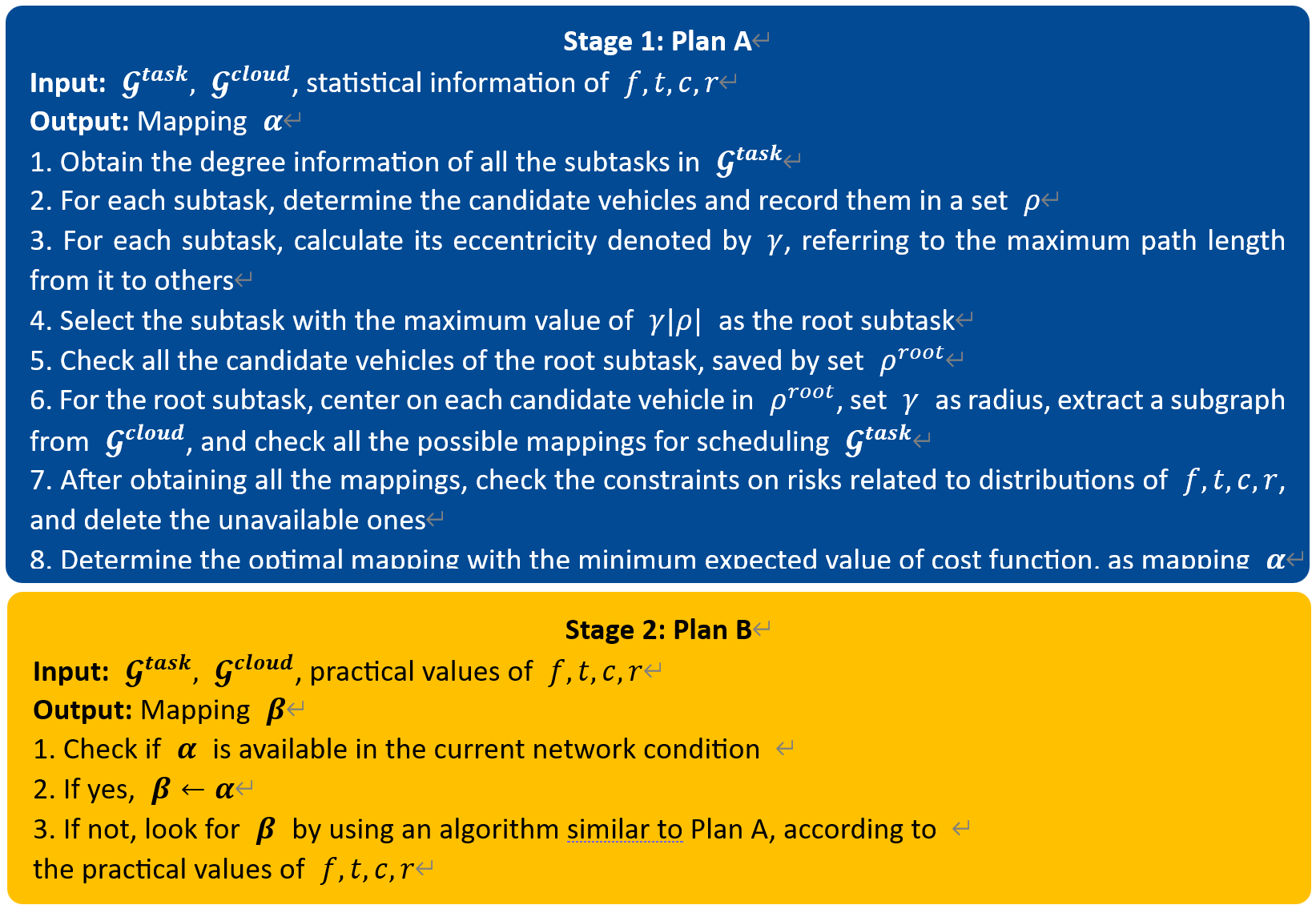}
\caption{Outline of our case study.}
\end{figure}

\subsection{Evaluation}
We compare the performance of our proposed \textbf{s}tage-\textbf{w}ise gr\textbf{a}ph \textbf{t}ask \textbf{s}cheduling mechanism (SWATS) against multiple baselines~\cite{2,15}: \textit{i)} onsite task scheduling (Onsite), which only uses Plan B; \textit{ii)} random task scheduling (Random), which randomly assigns subtasks to available vehicles at each task scheduling event~\cite{2}; \textit{iii)} time-preferred (TimePref) and degree-preferred (DegreePref) task scheduling, where TimePref maps each subtask to the vehicle with the lowest execution time, where DegreePref assigns each subtask to the vehicle with the largest number of V2V connections with others; and \textit{iv)} exhaustive search (ExSearch), which examines all possible mappings and selects the best one with the lowest cost function value at each practical scheduling event. We assume: $f$ (GHz) obeys a gaussian distribution with its mean in $[2,4]$ and variance in $[0.04,0.07]$ for each vehicle, $t$ (second) follows an exponential distribution with its mean falling in $[5,16]$, $c$ follows a normal distribution with its mean in $[0.03,0.07]$ and variance of 0.001, while $r$ (Mb/s) obeys a gaussian distribution with its mean in $[5,7]$ and variance of 0.55. Note that practical values of the above variables can neither be negative nor be too large in real-world networks due to the vehicular hardware settings and communication standards. We thus constrain/clip them as: $f\in[1.5,4.5]~\text{GHz}$, $t\in[0,60]~\text{seconds}$, $c\in[0.025,0.075]$ and $r\in[4,8]~\text{Mb/s}$~\cite{6,7}. We set the
weight coefficients in cost function as 0.5 since the task completion time and data exchange cost are on the same order of magnitude, according to our parameter settings. 
\vfill

In Fig. 5, we compare the average running time (ART) between our SWATS and other methods, which reflects the time spent on practical task scheduling events on searching for the best mappings. We consider 100 simulations (100 task scheduling events) to evaluate ART, where our results demonstrate the time effectiveness of SWATS.  
We use logarithmic representation to evaluate different methods for three task types given by Fig. 1 
As anticipated, the duration for each method to acquire feasible mappings increases with the larger number of subtasks and vehicles/connections, as indicated by the rising ART values. ExSearch has a particularly sharp increase in ART due to its large search space, which is not suitable for real-world dynamic IoV. Onsite method outperforms ExSearch by using a less complex approach, but its ART is still higher than others because it seeks to find all feasible mappings and chooses the best one. Although Random, TimePref, and DegreePref methods have lower ART values and sometimes perform better than SWATS, we will later show that they incur large costs. Our SWATS has commendable performance on ART because the prior obtained mapping $\alpha$ can be used in most practical scheduling events, making it suitable for dynamic MVCs. In Fig. 5, for tadpole task (type 3) with 15 vehicles, the value of ART can exceed $10^3$ seconds when using ExSearch, which is impractical. 

\begin{figure}[h!t]
\centering
\includegraphics[width=1\linewidth]{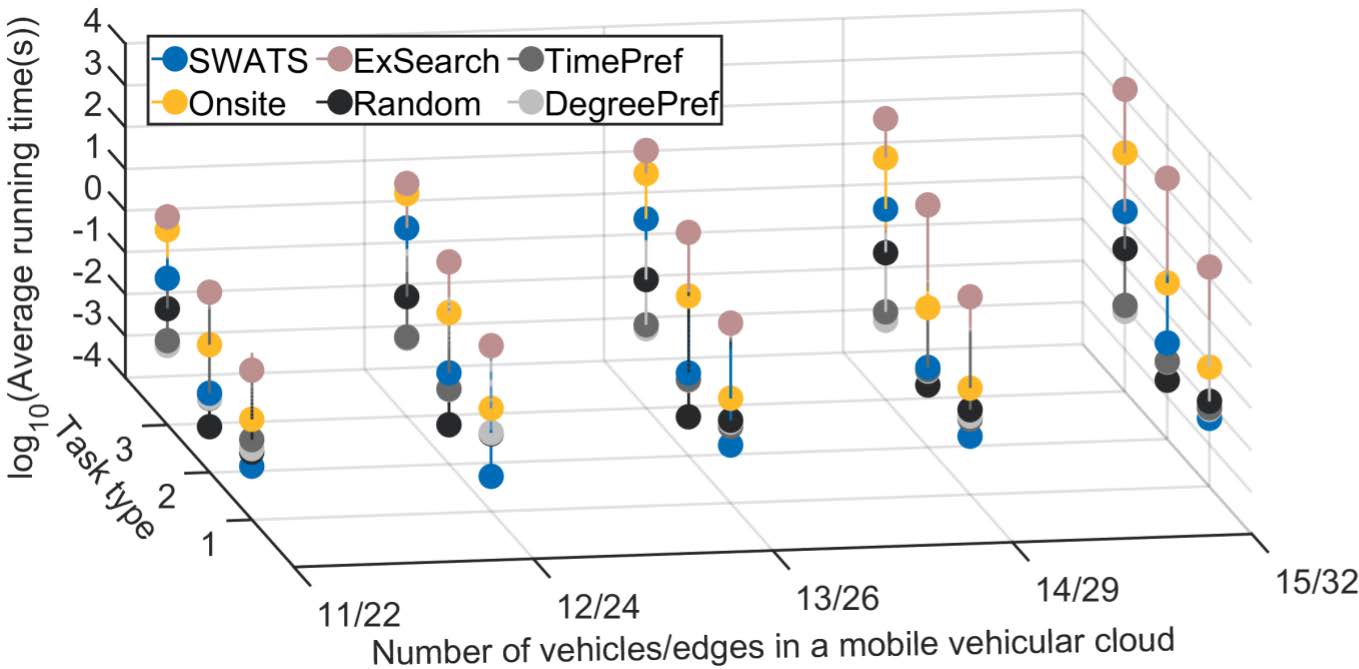}
\caption{Performance comparison in terms of the average running time.}
\end{figure}

\begin{figure}[h!t]
\centering
\includegraphics[width=1\linewidth]{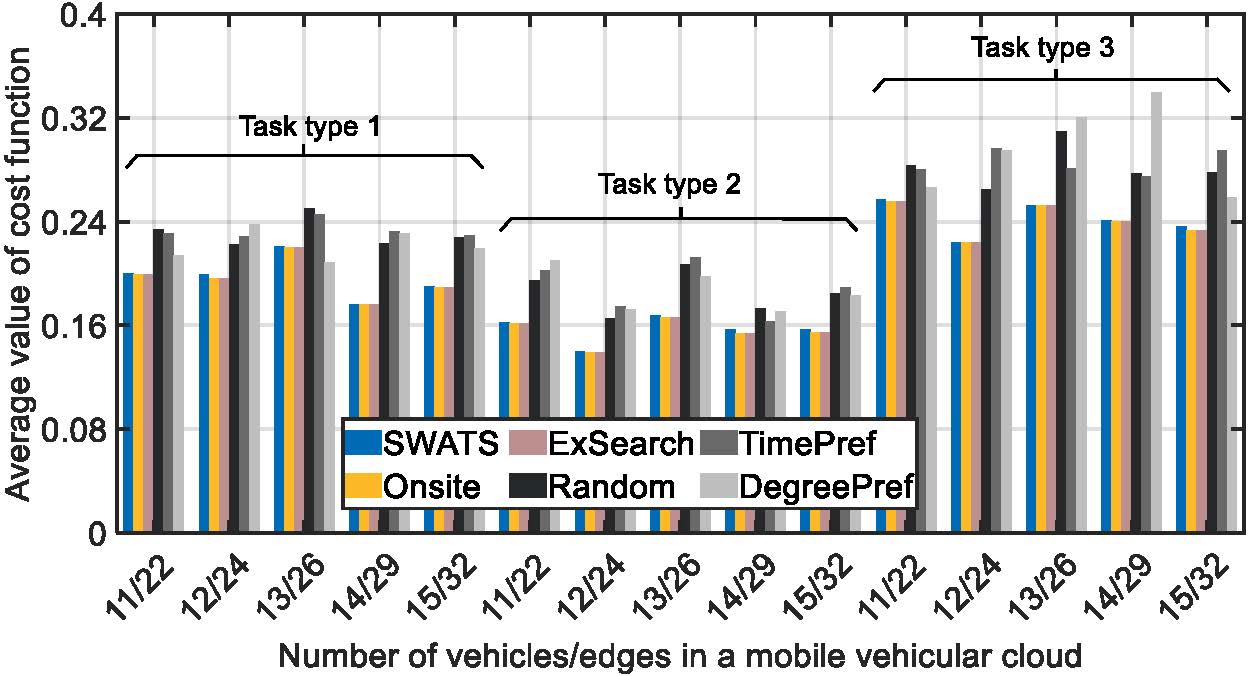}
\caption{Performance comparison in terms of the average value of cost function considering different task types.}
\end{figure}

%

We next depict the performance of SWATS, in terms of the average value of cost function (AVCF) and compare that to other methods in Fig. 6 for different task topologies. The results are consistent and reveal that SWATS greatly outperforms random and greedy-based methods due to the complementary mappings $\alpha$ and $\beta$. Onsite and ExSearch methods achieve the same value of AVCF since they both obtain all the possible mappings, while ExSearch suffers from unacceptable time overhead as shown in Fig. 5. Interestingly, SWATS performs similarly to the Onsite method on AVCF, while outperforming it on the value of ART (see Fig. 5), which reveals the importance of exploiting mapping $\alpha$ to achieve a high cost efficiency.

Overall, SWATS achieves a commendable performance on key evaluation indicators in comparison with existing methods, making the stage-wise decision-making paradigm a worthy reference for resource provisioning in future MVCs. 

\section{Future Directions}
In the following, we discuss several avenues of research.
\begin{itemize}[leftmargin=4mm] 
 \item \textit{Incentive design:} 
Considering resource provisioning over vehicles, a major issue is engaging them in the transmission, storage, and processing of data. One potential solution involves providing financial incentives to vehicles in exchange for their services. However, developing effective incentive mechanisms will require establishing a connection between resource provisioning methods, trading markets, and graph matching. Additionally, it will impose+ tackling nuanced challenges, e.g., establishing service prices to ensure maintaining the inherent task structure.

\item \textit{Overbooking-promoted data duplication:} 
Mobile vehicular networks exhibit unpredictability of fluctuations in resource supply and potential V2V outages. An interesting concept of ``overbooking" can be introduced, wherein each subtask can be assigned to more than one vehicle to account for potential service failures. However, overbooking may introduce excessive overhead, encompassing delays, increased energy consumption, and the risk of resource abuse. Consequently, optimizing the overbooking procedure can be considered as a pivotal focus to ensure high-performance task scheduling.

\item \textit{Accurate and timely environment assessment:} 
In wireless networks, uncertainties such as resource availability, communication disruptions, and the willingness of service providers are common occurrences. Inaccurate predictions of these uncertainties can have adverse effects on task scheduling performance. Hence, it is crucial to employ artificial intelligent (AI)-driven algorithms, such as time series prediction and recurrent neural networks, capable of swiftly and accurately acquiring information about the uncertainties in the environment.

\item \textit{Intelligent risk prediction, quantization, and management:} 
Our Plan A aims to uphold the promptness and cost-effectiveness of task scheduling by effectively balancing risks and opportunities. This calls for developing intelligent mechanisms for predicting, quantifying, and managing risks. Departing from traditional probabilistic models, the proposed mechanisms can leverage learning-based methods to model potential risks. Thus, future endeavors could explore establishing a connection between the domains of explainable artificial intelligence (XAI) and risk analysis to enhance the comprehensibility and interpretability of risk assessment.

\item \textit{Competition and cooperation among vehicles:} 
As computing service providers, vehicles can demonstrate a dynamic interplay of competition and cooperation. An intriguing research direction entails delving into mechanisms that promote constructive cooperation and revenue sharing among vehicles, all while taking into account the inherent structural complexities of graph tasks. This exploration carves out a novel research domain on executing graph tasks through cooperative games. Moreover, an alternative perspective on vehicle competition can be explored within the realm of auction theory. 

\item \textit{Addressing the curse of dimensionality:}
To reduce the problem scale, one way is to look for similarities between subtasks within a large graph task. For instance, we can group subtasks with similar attributes to form clusters. Another option is to group vehicles with similar resources and contact durations, treating them as a single unit. This can accelerate the optimization process by reducing the problem size.
\end{itemize}

\section{Conclusion}
We studied the allocation of computation-intensive tasks, modelled as undirected weighted graphs, over mobile vehicles within mobile vehicular clouds. To tackle the challenges of on-site scheduling in terms of excessive decision-making overhead, we introduced a novel stage-wise decision-making mechanism that encompasses two complementary plans for task scheduling: Plan A and Plan B. The objective of Plan A is to analyze historical statistics of uncertain factors like vehicle mobility and channel quality variations to obtain the best mapping $\alpha$ between subtasks and vehicles ahead of future task scheduling events. During each practical task scheduling event, Plan B is considered as a backup option to quickly identify a suitable mapping $\beta$ in case $\alpha$ fails. We investigated the procedure of our proposed paradigm and primary factors that contribute to its success, followed by a case study that includes simulations showcasing the performance evaluation in terms of cost and time efficiency. We also highlighted a set of research directions to further enhance resource provisioning over mobile vehicular clouds. 


\ifCLASSOPTIONcaptionsoff
 \newpage
\fi

\section*{Biography}
\vspace{-1cm}
\begin{IEEEbiographynophoto}{Minghui Liwang} is an assistant professor in School of Informatics, Xiamen University, China.
\end{IEEEbiographynophoto}
\vspace{-1cm}
\begin{IEEEbiographynophoto}{Bingshuo Guo} a Master student in Communication Engineering, Xiamen University, 
China. 
\end{IEEEbiographynophoto}
\vspace{-1cm}
\begin{IEEEbiographynophoto}{Zhanxi Ma} is a Ph.D student in School of Electronic and Engineering, Nanjing University, China.
\end{IEEEbiographynophoto}
\vspace{-1cm}
\begin{IEEEbiographynophoto}{Yuhan Su} is an assistant professor in School of Electronic Science and Engineering, Xiamen University, China. 
\end{IEEEbiographynophoto}
\vspace{-1cm}
\begin{IEEEbiographynophoto}{Jian Jin} is with Research Institute of Industrial Internet of Things, China Academy of Information and Communications Technology, China.
\end{IEEEbiographynophoto}
\vspace{-1cm}
\begin{IEEEbiographynophoto}{Seyyedali Hosseinalipour} is an assistant professor in Department of Electrical Engineering at University at Buffalo, SUNY.
\end{IEEEbiographynophoto}
\vspace{-1cm}
\begin{IEEEbiographynophoto}{Xianbin Wang} is a professor in Department of Electrical and Computer Engineering, Western University, Canada.
\end{IEEEbiographynophoto}
\vspace{-1cm}
\begin{IEEEbiographynophoto}{Huaiyu Dai} is a
professor in Department of Electrical and Computer Engineering, NC State University, USA.
\end{IEEEbiographynophoto}
\vfill

\begin{thebibliography}{15}



\bibitem{1}G. Panek, et al., “Application Relocation in an Edge-Enabled 5G System: Use Cases, Architecture, and Challenges,”~\textit{IEEE Commun. Mag.}, vol. 60, no. 8, pp. 28-34, 2022.

\bibitem{2} M. Liwang, et al., “Graph-Represented Computation-Intensive Task Scheduling Over Air-Ground Integrated Vehicular Networks,” \textit{IEEE Trans. Services Comput.}, vol. 16, no. 5, pp. 3397-3411, 2023.

\bibitem{3}W. Fan, et al., “Joint Task Offloading and Resource Allocation for Vehicular Edge Computing Based on V2I and V2V Modes,” \textit{IEEE Trans. Intell. Transp. Syst.}, vol. 24, no. 4, pp. 4277-4292, 2023.


\bibitem{4} Y. Wu, et al., “TDTA: Topology-Based Real-Time DAG Task Allocation on Identical Multiprocessor Platforms,” ~\textit{IEEE Trans. Parallel  Distrib. Syst.}, vol. 34, no. 11, pp. 2895-2909, 2023.

\bibitem{5}  J. Liu, et al., “A Proactive Stable Scheme for Vehicular Collaborative Edge Computing,” \textit{IEEE Trans. Veh. Technol.}, pp. 1--1, 2023.

\bibitem{6} Z. Gao, et al., “A Truthful Auction for Graph Job Allocation in Vehicular Cloud-Assisted Networks,” \textit{IEEE Trans. Mobile Comput.}, vol. 21, no. 10, pp. 3455--3469, 2022.

\bibitem{7} H. Zhu, et al., “Impact of Traffic Influxes: Revealing Exponential Intercontact Time in Urban VANETs,”~\textit{IEEE Trans. Parallel  Distrib. Syst.}, vol. 22, no. 8, pp. 1258-1266, 2011.

\bibitem{8} S. Hosseinalipour, et al., “Power-aware Allocation of Graph Jobs in Geo-Distributed Cloud Networks,” \textit{IEEE Trans. Parallel Distrib. Syst.}, vol. 31, no. 4, pp. 749--765, Apr. 2020.

\bibitem{9} H. Liao, et al., “Dependency-Aware Application Assigning and Scheduling in Edge Computing,”~\textit{IEEE Internet of Things J.}, vol. 9, no. 6, pp. 4451-4463, 2022.

\bibitem{10} S. Wi, S. Woo, J. J. Whang, and S. Son, “HiddenCPG: Large-Scale Vulnerable Clone Detection Using Subgraph Isomorphism of Code Property Graphs,”\textit{Proc. ACM Web Conf.}, Lyon, France, 2022, pp. 755--766.

\bibitem{11}W. Feng, et al., “Latency Minimization of Reverse Offloading in Vehicular Edge Computing,”\textit{IEEE Trans. Veh. Technol.}, vol. 71, no. 5, pp. 5343-5357, 2022.

\bibitem{12}  S. Sheng, et al., “Futures-based Resource Trading and Fair Pricing in Real-Time IoT Networks,” \textit{IEEE Wireless Commun. Lett.}, vol. 9, no. 1, pp. 125--128, 2020.

\bibitem{13} M. Abulaish, et al., “Subiso: A Scalable and Novel Approach for Subgraph Isomorphism Search in Large Graph,” \textit{IEEE Int. Conf. Commun. Syst. Netw.}, Bengaluru, India, pp. 102--109, 2019.

\bibitem{14} Y. Liu, et al., “Secure and Efficient Stigmergy-Empowered Blockchain Framework for Heterogeneous Collaborative Services in the Internet of Vehicles,”~\textit{IEEE Commun. Mag.}, vol. 61, no. 9, pp. 186-192, 2023.

\bibitem{15} S. Luo, et al., “HFEL: Joint Edge Association and Resource Allocation for Cost-Efficient Hierarchical Federated Edge Learning,”~\textit{IEEE Trans. Wireless Commun.}, vol. 19, no. 10, pp. 6535-6548, 2020.
\end{thebibliography}
\end{document}